# Long triple carbon chains formation by heat treatment of graphene nanoribbon: molecular dynamics study with revised Brenner potential

Alexander S. Sinitsa[a], Irina V. Lebedeva[b], Andrey M. Popov[c*], and Andrey A. Knizhnik[d]

[a] National Research Centre "Kurchatov Institute", Kurchatov Square 1, Moscow 123182, Russia.

[b] Nano-Bio Spectroscopy Group and ETSF, Universidad del País Vasco, CFM CSIC-UPV/EHU, San Sebastian 20018, Spain.

[c] Institute for Spectroscopy of Russian Academy of Sciences, Fizicheskaya Street 5, Troitsk, Moscow 108840, Russia.

[d] Kintech Lab Ltd., 3rd Khoroshevskaya Street 12, Moscow 123298, Russia.

**Abstract**

The method for production of atomic chains by heating of graphene nanoribbons (GNRs) is proposed and studied by molecular dynamics simulations. The Brenner potential is revised to adequately describe formation of atomic chains, edges and vacancy migration in graphene. A fundamentally different behavior is observed for zigzag-edge GNRs with 3 and 4 atomic rows (3 and 4-ZGNRs) at 2500 K: formation of triple, double and single carbon chains with the length of hundreds of atoms in 3-ZGNRs and edge reconstruction with only short chains and GNR width reduction in 4-ZGNRs. The chain formation mechanism in 3-ZGNRs is revealed by analysis of bond reorganization reactions and is based on the interplay of two processes. The first one is breaking of bonds between 3 zigzag atomic rows leading to triple chain formation. The second one is bond breaking within the same zigzag atomic row, which occurs predominantly through generation of pentagons with subsequent bond breaking in pentagons and results in single or double chain formation. The DFT calculations of the barriers for relevant reactions are consistent with the mechanism proposed. The possibility of chain-based nanoelectronic devices with a controllable number of chains is discussed.

## 1. Introduction

Carbon atomic chains, sp-bonded linear allotrope of carbon, have been studied intensively during the last decades and demonstrate unusual chemical, mechanical, optical, electronic and transport properties [1-15]. Metal-decorated carbon chains have been proposed for hydrogen storage [12]. Calculated transport properties of carbon chains show that they hold much promise for application in electronic [7,11,14,15] and spintronic [4,8,9,15] devices. According to the recent density functional theory (DFT) calculations, semiconductor-metal transition and spin polarization occur in different ways in nanodevices with single, double and triple carbon chains between electrodes [15]. Elaboration of methods to fabricate carbon chain-based nanodevices with a controllable number of chains hold the key to success of their applications.

---

[*] Corresponding author. Tel. +7-909-967-2886 E-mail: popov-isan@mail.ru (Andrey Popov)

Whereas methods of mass carbon chain synthesis in the gas phase have been elaborated (see, for example Ref. 16), they only allow to produce films containing sp carbon chains [17]. Carbon chains have been also obtained by various ways inside carbon nanotubes [18-20]. Chemical methods for synthesis of carbon chains up to 44 atoms in length [21] should also be mentioned. However, these developments have not yet made possible fabrication of nanodevices based on individual carbon chains. The success in this field has been achieved by elaboration of top-down methods to produce carbon chains by etching of graphene nanoribbons (GNRs) [22-25] and carbon nanotubes [26,27] by electron irradiation [22,23,25,26,28], and by combined etching using electron irradiation and Joule heat [24]. Not only such etching has allowed to obtain single, double [22,23,28], and triple [23,28] chains but also carbon chain-based electronic nanodevices have been implemented and conductance of carbon chain has been measured [24]. The maximal chain length of about 60 atoms has been reached by the combined action of electron irradiation and Joule heat on a GNR. For further progress of the top-down methods of carbon chain production, it is necessary to choose the appropriate initial nanoobject and the way of treatment of this nanoobject that causes efficient formation of the chains. A number of different narrow GNRs have been synthesized recently [28-33] and we believe that such GNRs can be candidates for initial nanoobjects for chain production. Moreover, GNR transfer from the substrate where the GNR is synthesized to a nanodevice before the treatment is possible [34]. Here we propose synthesis of long carbon chains under heat treatment of narrow GNRs and confirm this suggestion by molecular dynamics (MD) simulations.

Both the bottom-up process of chain formation as a result of $C_{10}H_2$ molecules fusion [20] and top-down chain formation by rupture of bulk graphene [41,43] and GNRs [38-40,42,44], evaporation of GNRs at high temperature [35,36], and as a result of GNR etching by electron irradiation [37] have been studied by *ab initio* and classical atomistic simulations. In these simulations, the longest chains consisting of 15-20 atoms have been observed for graphene rupture along grain boundaries [43] and at high temperature [44]. The chains with the length of only up to 10 atoms are formed during the rupture of pristine graphene at room temperature [38-42]. The atomistic mechanism of formation of long chains in these cases can be associated with inhomogeneity in stress distribution introduced by pentagons and heptagons existing at grain boundaries [43] or arising at high temperature [44]. Classical MD simulations performed here reveal a principally different atomistic mechanism of chain formation and predict formation of triple, double and single carbon chains of up to hundreds of atoms in length during heating of zigzag-edged GNRs with 3 atomic rows (3-ZGNR). On the other hand, we show that heating of 4-ZGNR leads to formation only of short chains [38-44] (the nomenclature of GNRs from Ref. 45 is used).

MD simulations based on empirical potentials are widely used to get insight into properties of systems of a large size and processes taking a long time, which are not accessible for MD simulations based on tight-binding potentials and *ab initio* MD. Nevertheless, care should always be taken when one is interested in the properties beyond the potential training set. A number of complex semiempirical potentials able to describe bond formation and dissociation are available for carbon: Brenner reactive bond-order potentials (REBO) [46,47], ReaxFF [48] and LCBOPII [49], etc. All the potentials listed are fitted to the properties of small hydrocarbons and regular carbon crystals. Though ReaxFF was extended recently to describe amorphous carbon clusters and defects in graphene [50], hydrocarbons were still an important part of the training set. Such wide-scope potentials represent a good choice for qualitative description of an arbitrary hydrocarbon system. However, their accuracy for some properties of carbon nanostructures like graphene, carbon nanotubes and fullerenes, which are large regular systems but with irregularities in the form of edges and defects, is not satisfactory. In some cases the potentials are not even qualitatively correct. It has been demonstrated that the potentials listed fail to reproduce the relative stability of different graphene edges [51,52,53]. The latest versions of the REBO potential also give the wrong ground-state structure for vacancies in graphene [54].

We show that in the case of the REBO potentials, the deficiencies critical for simulations of transformations of carbon nanostructures can be easily fixed by adjustment of several parameters, which were originally derived from atomization energies of small hydrocarbons and are not necessarily the same for carbon nanostructures. Though in this way we restrict the area of applicability of the potentials to carbon nanostructures, the description of these particular systems is improved. In addition to energetics of graphene edges and vacancy migration in graphene, we also fit the formation energy of atomic carbon chains, which often arise in carbon nanostructures at high temperature or under electron irradiation (see, e.g., papers on transformation of graphene flakes [51,55-57] and amorphous carbon clusters [52] to fullerenes and reorganization of carbon nanotube structure upon cutting with a metal cluster [58]). The REBO-1990 potential [46] with the modified set of parameters is applied here for simulations of chain formation from GNRs. The adequacy of the potential is confirmed by the comparison of the barriers of principal reactions leading to chain formation with the results of DFT calculations.

In GNRs edges have a crucial effect on the physical properties and chemical reactivity [59]. Under-coordinated atoms at the edges are already destabilized compared to those inside graphene layers and get involved into structural transformations more easily. Nevertheless, reactions at the edges are much less studied compared to rearrangements within the layers. Available DFT data on barriers of such processes for purely carbon systems are limited to the cases of simultaneous and complete reconstruction of the zigzag edge [53,60,61] and formation of the first [25,62,63] and second [62] pentagon-heptagon pairs. Formation of atomic carbon chains at the graphene edges has been so far considered only using empirical potentials [56,57,64]. We present the results of DFT calculations on consecutive bond breaking at GNR edges leading to generation of carbon atomic chains of the length of up to 5 atoms. Furthermore, we compare the potential energy profiles for formation of carbon atomic chains and generation of pentagon-heptagon pairs in 3 and 4-ZGNRs.

The paper is organized the following way. Section 2 is devoted to fitting of the REBO potentials to the properties related to chain formation and details of MD and DFT calculations. Section 3 presents the results of MD simulations of chain formation and DFT calculations of the barriers for the most important reactions during this process. A scheme of a chain-based nanoelectronic device is also proposed. Our conclusions are summarized in Section 4.

## 2. Methods
*2.1 Modification of interatomic potential*
*2.1.1. Description of REBO potentials*
In REBO potentials [46, 47], the binding energy is represented as a sum over bonds

$$E_b = \sum_i \sum_{j>i} \left( V_R(r_{ij}) - \bar{B}_{ij} V_A(r_{ij}) \right), \qquad (1)$$

where $V_R(r_{ij})$ and $V_A(r_{ij})$ are repulsive and attractive pair terms for atoms $i$ and $j$ at distance $r_{ij}$ and $\bar{B}_{ij}$ is the bond-order term that allows to describe the dependence of the bond energy on the coordination of the atoms. In the following we consider purely carbon systems and all terms with hydrogen are omitted.

The bond-order term is given by

$$\bar{B}_{ij} = \left( B_{ij} + B_{ji} \right)/2 + B_{\mathrm{DH}} + F\left( N_{ij}^t, N_{ji}^t, N_{ij}^{conj} \right)/2, \qquad (2)$$

where $B_{ij}$ and $B_{ji}$ are bond orders for atoms $i$ and $j$, $B_{\mathrm{DH}}$ describes the dependence on the dihedral angles for REBO-2002 [47] (this term is zero for REBO-1990 [46]) and $F$ is the correction function depending on the total numbers of neighbours $N_{ij}^t$ and $N_{ji}^t$ for atoms $i$ and $j$ except for themselves and on whether the bond is a part of the conjugated system or not according to the value of $N_{ij}^{conj}$.

The total number of neighbours is computed as

$$N_{ij}^t = \sum_{k \neq i,j} f(r_{ik}), \qquad (3)$$

where $f(r)$ is the cutoff function equal to 1 for close distances and 0 for large ones:

$$f(r) = \begin{cases} 1, & r < R^{(1)} \\ \frac{1}{2}\left(1 + \cos\frac{\pi(r - R^{(1)})}{(R^{(2)} - R^{(1)})}\right), & R^{(1)} \leq r \leq R^{(2)} \\ 0, & r > R^{(2)} \end{cases} \quad (4)$$

In REBO-1990 [46], $N_{ij}^{conj}$ is given by

$$N_{ij}^{conj} = 1 + \sum_{k \neq i,j} f(r_{ik}) \Phi(N_{ki}^t) + \sum_{k \neq i,j} f(r_{jk}) \Phi(N_{kj}^t), \quad (5)$$

where $\Phi(x)$ changes from 1 for $x \leq 2$ to 0 for $x \geq 3$:

$$\Phi(x) = \begin{cases} 1, & x \leq 2 \\ (1 + \cos\pi(x - 2))/2, & 2 < x < 3 \\ 0, & x \geq 3 \end{cases} \quad (6)$$

In REBO-2002, a slightly different expression is used

$$N_{ij}^{conj} = 1 + \left[\sum_{k \neq i,j} f(r_{ik}) \Phi(N_{ki}^t)\right]^2 + \left[\sum_{k \neq i,j} f(r_{jk}) \Phi(N_{kj}^t)\right]^2. \quad (7)$$

The correction $F$ is needed to describe bond breaking and formation [46,47]. If the function $F$ in eq. (2) is omitted, the simple average is used for the bond orders of the atoms. Let us consider the situation when one of the bonded atoms is three-coordinated and another is four-coordinated. Without the function $F$, the bond between them would have a mixed single and double-bond character. The real physical picture, however, is that the system is described as a single bond and a radical. This is solved through dependence of the function $F$ on the numbers of neighbours for each atom in the bond. Another role of the function $F$ is to distinguish between conjugated and non-conjugated bonds.

The parameters of the pairwise terms and bond orders for each atom in the REBO potentials were fitted to the experimental and *ab initio* data on lattice constants and cohesive energies of regular carbon crystals: graphite, diamond, simple cubic and face-centered cubic structures of carbon. In REBO-2002, the force constants of graphite and diamond were also taken into account. The values of the function $F(n_1, n_2, n_3)$ for discrete numbers $n_1, n_2$ and $n_3$ were derived from atomization energies of small hydrocarbons and tight-binding results on formation energy of vacancies in graphite and diamond. The values of the function $F$ for non-integer numbers $n_1, n_2$ and $n_3$ are obtained by interpolation. It should be noted that $F(n_1, n_2, n_3) = F(n_2, n_1, n_3)$ and

$F(n_1, n_2, n_3 > n_0) = F(n_2, n_1, n_0)$, where $n_0 = 2$ in REBO-1990 and 9 in REBO-2002. We consider only the second set of parameters for the REBO-1990 potential (Table III of Ref. 46), which describes better the force constants.

*2.1.2. Fitting procedure*

Inside the graphene layer, $F(2,2,2) = 0$ in REBO-1990 and $F(2,2,9) = 0$ in REBO-2002 and we do not touch these values. However, we admit that other values of the function $F$ can be modified for carbon nanostructures since they were fitted to the properties of small hydrocarbons. The absolute formation energy of vacancies in graphite fitted using the parameters $F(1,2,2)$ in REBO-1990 and $F(1,2,9)$ in REBO-2002 is also very large (about 7 eV) and formation of vacancies is unlikely to occur at high temperature or under irradiation by electrons with a moderate kinetic energy. Therefore, we assume that it is sufficient to describe it with the accuracy of 1 eV. On the other hand, migration of vacancies, reconstruction of edges of graphene and formation of carbon atomic chains at the edges can take place under the conditions of interest and we find it important to describe these properties properly.

The parameters of REBO-1990 and REBO-2002 are changed in three consecutive steps to reproduce (1) the energies of graphene edges, (2) the energy of the symmetric saddle point for vacancy migration in graphene relative to the ground-state vacancy structure and finally (3) the formation energy of atomic carbon chains. Though in the present paper we first perform step (1) and then (2), these two steps are actually independent as the energies of graphene edges and the relative energy of the symmetric saddle point for vacancy migration are determined by values of the function $F$ for different coordination numbers of carbon atoms. The corresponding versions of the potentials are marked by letters ´E´ and ´V´ (or ´EV´ for both of the steps). For the step (3) the parameters corresponding to the steps (1) and (2) have to be readjusted to describe all the characteristics (1) – (3). The letter ´C´ in the name of the potential version indicates that the formation energy of atomic carbon chains is also fitted.

Along with the REBO potentials, we also test the ReaxFF [48] force field for condensed carbon phases, which was recently fitted (ReaxFF$_{C-2013}$) to the experimental data on heats of formation of hydrocarbons and carbon crystals and the results of DFT calculations for equations of state for diamond and graphite, formation energies of defects in graphene and heats of formation of amorphous carbon clusters [50]. Another popular empirical potential for simulations of purely carbon systems is LCBOPII [49]. It was developed on the basis of the REBO potentials with the aim of description of various liquid and solid carbon phases with account of non-bonded interactions. The training set for the short-range part of this potential was similar to the one of the REBO potentials. Furthermore,

LCBOPII was recently modified to correct the description of graphene edges and we cite the corresponding data from Ref. 53.

*2.1.3. First step: graphene edges*

As described above, at the first step we revise the energy of graphene edges. Three types of graphene edges are considered: armchair (AC), zigzag (ZZ) and reconstructed zigzag (ZZ(57)), which is formed from the ZZ edge by transformation of pairs of adjacent hexagons to pentagons and heptagons. Numerous spin-polarized DFT calculations [60,61,65–71] (see also review [59]) have been performed to study these edges using the local density approximation (LDA) [72] and the generalized gradient approximation with the exchange-correlation functionals of Perdew, Burke and Ernzerhof (PBE) [73], of Perdew and Wang (PW91) [74] and others. It is known from these studies that the reconstruction of the ZZ edge reduces the edge energy by 0.2–0.4 eV/Å (Refs. 60, 61, 65–71, Table 1). Such a reconstructed edge is even slightly more stable than the AC edge and it is often observed experimentally [75-78]. The close energies of the ZZ(57) and AC edges are related to similarity in their geometries. While two-coordinated atoms in the unreconstructed ZZ edge have dangling bonds responsible, among others, for significant spin polarization, two-coordinated atoms in ZZ(57) and AC edges form triple bonds of 1.22–1.26 Å length [61,65,66,69,71] (close to the 1.20 Å triple bond in acetylene) and the absence of dangling bonds is manifested through the loss of spin polarization. Taking into account the scatter in the DFT data, we assume that the potentials are fitted well if the difference in the energies for ZZ and ZZ(57) edges lies in the 0.2–0.3 eV/Å range and the difference in the energies for AC and ZZ(57) edges is positive and below 0.1 eV/Å. The absolute energy of the ZZ(57) should lie from 0.95 eV to 1.1 eV.

To calculate the energies of graphene edges we have performed geometrical optimization of an AGNR and ZGNRs with pristine and reconstructed edges. Details of these calculations can be found below in Section 2.2. The edge energies per unit edge length are computed as $E_{ed} = (E_{GNR} - N_{GNR}\varepsilon_{gr})/L_{ed}$, where $E_{GNR}$ is the ribbon energy, $N_{GNR}$ is the number of atoms in the ribbon per unit cell, $\varepsilon_{gr}$ is the energy of bulk graphene per atom and $L_{ed}$ is the total edge length per unit cell. These calculations show that the original REBO-1990 potential somewhat underestimates the edge energies, though qualitatively the stability of different edges is well described (Table 1). REBO-2002 and ReaxFF$_{C-2013}$ fail even to describe the correct order of graphene edge energies assigning erroneously the lowest energy to the unreconstructed ZZ edge. The original LCBOPII gives almost the same edge energies for all three ZZ, ZZ(57) and AC edges [53]. In the modified version, the ZZ(57) and AC edges are stabilized relative to the ZZ edge but the absolute edge energies are quite strongly underestimated. For the both versions of

LCBOPII, the most stable edge is AC and not ZZ(57), as follows from the DFT calculations.

The energies of ZZ, ZZ(57) and AC edges are determined by $F(1,2,2)$ and $F(1,1,2)$ for REBO-1990 and $F(1,2,9)$ and $F(1,1,9)$ for REBO-2002. Note that in original REBO-2002, $F(1,1,n_3)$ is the same for $n_3 = 3-9$ and $F(1,2,n_3)$ is the same for $n_3 = 6-9$ and we maintain this property. The modified REBO-1990 potential with the parameter $F(1,2,2) = -0.063$, as proposed in our previous paper [51], improves the magnitudes of the graphene edge energies as compared to the original version [46] of the potential (Table 1). However, even more justified values are obtained when we set this parameter to -0.053 in REBO-1990E at the first step of revision of the potential. In the case of REBO-2002 we have not succeeded in adjusting the parameters so that the reconstructed ZZ edge would be more stable than the AC one but at least it is possible to destabilize the ZZ edge and to provide close energies for the reconstructed ZZ and AC edges (Table 1). Destabilization of the ZZ edge, however, is accompanied by the increase in the absolute energy of the ZZ(57) edge. Trying to balance description of these two properties, neither the absolute energy of the ZZ(57) edge nor the relative energy of the ZZ edge relative to ZZ(57) get into desired intervals.

*2.1.4. Second step: vacancy migration*
At the second step of revision of the potentials we consider formation and migration of a monovacancy in graphene. Though removal of one carbon atom in a perfect graphene layer leads to formation of the monovacancy with the $D_{3h}$ symmetry and three dangling bonds [79,80], the DFT calculations show that Jahn-Teller distortion breaks the local three-fold symmetry down to $C_{2v}$ (see Refs. 81–88 and review [59]). As a result, the reconstructed vacancy has a so-called 5/9 structure, where the new bond is formed between the pair atoms with dangling bonds giving rise to a pentagon and a nine-membered ring and only one dangling bond is left. Both the reconstructed [89-91] and symmetric structures [89] of the vacancy have been observed experimentally. The experimental estimate of the formation energy of monovacancies is $E_v = 7.0 \pm 0.5$ eV [92]. DFT calculations give results in the 7 − 8 eV range and some of them are listed in Table 2. It should be noted, however, that it is rather difficult to converge the formation energy of vacancies with respect to the size of the model cell in DFT calculations [59]. Therefore, it can be expected that the recent result of 7.4 eV from PBE calculations [82] for the model cell with $N_0 = 288$ atoms before vacancy formation is the most reliable and indeed it lies within the error bars of the experimental value. In simulations at high temperature or under irradiation by electrons with a moderate kinetic energy, it is not critical to describe the formation energy of vacancies with high precision and the values within the 6 – 8 eV are reasonable.

Table 1. Energies of the reconstructed zigzag ($E_{ZZ(57)}$), unreconstructed zigzag ($E_{ZZ}$) and armchair edges ($E_{AC}$) of graphene (in eV/Å) obtained using different interatomic potentials and by DFT calculations.

| Ref. | Ref. | Parameters | $E_{ZZ(57)}$ | $E_{AC}$ | $E_{ZZ}$ | $E_{AC} - E_{ZZ(57)}$ | $E_{ZZ} - E_{ZZ(57)}$ |
|---|---|---|---|---|---|---|---|
| This work | REBO-1990 (Ref. 46) | $F(1,2,2) = -0.0243$, $F(2,3,2) = -0.0363$, $F(1,1,2) = 0.0108$ | 0.937 | 1.000 | 1.035 | 0.063 | 0.098 |
| This work | Modified REBO-1990 (Ref. 51) | $F(1,2,2) = -0.063$, $F(2,3,2) = -0.0363$, $F(1,1,2) = 0.0108$ | 1.147 | 1.247 | 1.494 | 0.099 | 0.346 |
| This work | REBO-1990E/EV | $F(1,2,2) = -0.053$, $F(2,3,2) = -0.0363/-0.104$, $F(1,1,2) = 0.0108$ | 1.095 | 1.185 | 1.378 | 0.090 | 0.284 |
| This work | REBO-1990EVC | $F(1,2,2) = -0.038$, $F(2,3,2) = -0.088$, $F(1,1,2) = 0.02514818$ | 0.966 | 1.034 | 1.201 | 0.067 | 0.235 |
| This work | REBO-2002 (Ref. 47) | $F(1,1,3-9) = -0.0160856$, $F(1,2,6-9) = -0.030133632$, $F(2,3,9) = -0.044709383$ | 1.124 | 1.091 | 1.041 | −0.034 | −0.084 |
| This work | REBO-2002E/EV | $F(1,1,3-9) = 0.004077$, $F(1,2,6-9) = -0.063$, $F(2,3,9) = -0.044709383/-0.332$ | 1.211 | 1.189 | 1.351 | −0.021 | 0.140 |
| This work | ReaxFF$_{C-2013}$ (Ref. 50) |  | 1.110 | 1.207 | 1.089 | 0.097 | −0.021 |
| [53] | LCBOPII (Ref. 49) |  | 1.06 | 1.04 | 1.05 | −0.02 | −0.01 |
| [53] | Modified LCBOPII (Ref. 53) |  | 0.81 | 0.75 | 1.05 | −0.06 | 0.24 |
| [65] | DFT-LDA |  | 1.09 | 1.10 | 1.34 | 0.01 | 0.25 |
| [60] | DFT-LDA |  | 1.06 | 1.09 | 1.43 | 0.03 | 0.37 |
| [66] | DFT-LDA |  | 1.147 | 1.202 | 1.391 | 0.055 | 0.244 |
| [53] | DFT-PBE |  | 0.98 | 1.02 | 1.15 | 0.04 | 0.17 |
| [61] | DFT-PBE |  | 0.96 | 0.98 | 1.31 | 0.02 | 0.35 |
| [67] | DFT-PBE |  | 0.965 | 1.008 | 1.145 | 0.043 | 0.180 |
| [68] | DFT-PBE |  | 0.98 | 0.99 | 1.18 | 0.01 | 0.20 |
| [69] | DFT-PBE |  |  |  |  |  | 0.148 |
| [70] | DFT-PW91 |  | 0.97 |  | 1.21 |  | 0.24 |
| [71] | DFT-GGA |  | 0.97 | 1.00 | 1.17 | 0.03 | 0.20 |

Table 2. Formation energy $E_{5/9}$ of 5/9 vacancies in graphene and energy $\Delta E_{sym}$ of the symmetric saddle point for vacancy migration (planar spiro state) relative to the 5/9 ground state (in eV) obtained using interatomic potentials and by DFT calculations for model cells with $N_0$ atoms in the graphene layer before atom removal.

| Ref. | $N_0$ | Method | Parameters | $E_{5/9}$ | $\Delta E_{sym}$ |
|---|---|---|---|---|---|
| This work | 720 | REBO-1990 (Ref. 46) | $F(1,2,2) = -0.0243$, $F(2,3,2) = -0.0363$, $F(1,1,2) = 0.0108$ | 6.272 | 0.169 |
| This work | 720 | Modified REBO-1990 (Ref. 51) | $F(1,2,2) = -0.063$, $F(2,3,2) = -0.0363$, $F(1,1,2) = 0.0108$ | 7.380 | −2.065 |
| This work | 720 | REBO-1990E | $F(1,2,2) = -0.053$, $F(2,3,2) = -0.0363$, $F(1,1,2) = 0.0108$ | 7.114 | −0.673 |
| This work | 720 | REBO-1990EV | $F(1,2,2) = -0.053$, $F(2,3,2) = -0.104$, $F(1,1,2) = 0.0108$ | 7.114 | 1.253 |
| This work | 720 | REBO-1990EVC | $F(1,2,2) = -0.038$, $F(2,3,2) = -0.088$, $F(1,1,2) = 0.02514818$ | 6.692 | 1.232 |
| This work | 720 | REBO-2002 (Ref. 47) | $F(1,1,3\text{-}9) = -0.0160856$, $F(1,2,6\text{-}9) = -0.030133632$, $F(2,3,9) = -0.044709383$ | 6.962 | −2.005 |
| This work | 720 | REBO-2002E | $F(1,1,3\text{-}9) = 0.004077$, $F(1,2,6\text{-}9) = -0.063$, $F(2,3,9) = -0.044709383$ | 7.725 | −2.768 |
| This work | 720 | REBO-2002EV | $F(1,1,3\text{-}9) = 0.004077$, $F(1,2,6\text{-}9) = -0.063$, $F(2,3,9) = -0.332$ | 7.724 | 1.241 |
| This work | 720 | ReaxFF$_{C\text{-}2013}$ (Ref. 48) |  | 8.99 | −1.21 |
| [83] | 128 | DFT-PW91 |  | 7.73 | 1.33 |
| [84] | 32 | DFT-PW91 |  | 7.88 | 1.40[a], 0.99[b] |
| [85] | 128, 100[a] | DFT-PW91 |  | 7.85 | 1.37, 1.26[a] |

| [79] | 18 | PBE-LDA | | 7.6 [a,c] | 1.6 [a,c] |
|---|---|---|---|---|---|
| [80] | 128 | DFT-LDA | | | 1.01 [c] |
| [86,87] | 60 | DFT-PBE | | | 1.17 |
| [88] | 288 | DFT-LDA, DFT-PBE | | | 1.2 |
| [82] | 288 | DFT-PBE | | 7.36 | |
| | 288 | DFT-LDA | | 7.91 | |

[a] For graphite.
[b] For few-layer graphene.
[c] For unreconstructed vacancy.

As for the interatomic potentials considered in the present paper, all of them agree that vacancy reconstruction to the 5/9 structure is energetically favourable and the predicted formation energies of the 5/9 vacancy calculated as $E_{5/9} = E_d - (N_0 - 1)\varepsilon_{gr}$, where $E_d$ is the total energy of the structure with the vacancy, are mostly within the desired range (Table 2). The details of these calculations are discussed in Section 2.2. Nevertheless, it can be mentioned that while the original REBO-1990 somewhat underestimates the formation energy as compared to the experimental and most accurate DFT results, the modified version of this potential from Ref. 51 and REBO-1990E give the formation energies within the experimental range. REBO-2002 and REBO-2002E tend to slightly overestimate the formation energy and ReaxFF$_{C-2013}$ overestimates the formation energy by almost 2 eV.

A more important failure in the performance of the potentials is that REBO-1990E, REBO-2002, REBO-2002E and ReaxFF$_{C-2013}$ predict that another structure is more stable than the 5/9 vacancy (Table 2). This structure referred to in literature as "spiro" state [59] has a carbon atom at equal distances from four other atoms. The planar spiro state, which corresponds to the symmetric saddle point for vacancy migration, has been well characterized in DFT calculations (all the data known to us are listed in Table 2) and we use the energy $\Delta E_{sym}$ of the planar spiro state relative to the ground 5/9 state of the vacancy from these calculations as a benchmark for potential fitting. We assume that this property is described well if it is in the 1.0 – 1.4 eV range. It should be also noted that though the original REBO-1990 potential [46] predicts that the spiro state is slightly unstable compared to the ground 5/9 state, the energy difference for these two states is too small compared to the DFT calculations. Changing the parameters $F(2,3,2)$ and $F(2,3,9)$ in REBO-1990EV and REBO-2002EV, respectively, at the second step of revision of the potentials we adjust the relative energy of the planar spiro state to about 1.2 eV keeping the formation energy in the acceptable limits (Table 2).

Though the planar spiro state is often considered as the transition state for vacancy migration [83-88], recent DFT calculations [81] demonstrate that in reality the transition state, while still being spiro state with a carbon atom forming four equal bonds, has a non-planar structure. The corresponding PBE result for the barrier to vacancy migration is about 0.9 eV and it is very close to the values of 0.9–1.0 eV deduced experimentally from the direct observation of vacancy diffusion on a graphite surface with scanning tunneling microscopy [93]. The ReaxFF$_{C-2013}$ potential also shows the reduction of the spiro state energy by 0.09 eV upon taking into account atomic displacements out of plane. With the REBO potentials, however, the spiro state is optimized to the planar geometry even when initially some atoms are displaced out of plane. This is related to the symmetry of the system and is not fixed by a simple adjustment of several parameters in the potential. Nevertheless, the corrected relative energy of the planar spiro state in REBO-2002EV and REBO-1990EV already implies significant improvement in the description of vacancy dynamics in graphene.

*2.1.5. Third step: chain formation*

Another important characteristic for simulations at high temperature or under electron irradiation is the formation energy of atomic carbon chains and we include it into consideration at the third (final) step of revision of the potentials. The benchmark data for this energy have been obtained by spin-polarized DFT calculations as described in Section 2.3. According to our PBE (PW91) calculations, the optimized structures of the atomic chains have alternating bonds of 1.26 (1.25) and 1.31 (1.30) Å length, within the ranges of previous DFT calculations [1,9,13,14,18,22,24] and experimental data [27,94]. The formation energy of chains per atom is calculated as the difference in the energy per atom in the chains and graphene and is found to be 0.98 (0.95) eV according to the PBE (PW91) calculations, in agreement with the previous DFT results of 0.97 eV (Ref. 22) and about 1 eV (Ref. 27). We assume that potentials describe well the formation energy of chains if it lies in the range from 0.95 eV to 1 eV.

ReaxFF$_{C-2013}$ predicts the formation energy of chains per atom of 1.19 eV. The original REBO-1990 and REBO-2002 potentials give 1.20 eV and 1.30 eV, respectively. These quantities can be tuned by changing $F(1,1,2)$ for REBO-1990 and $F(1,1,3)$ for REBO-2002. Using the parameters $F(1,1,3-9)=0.004077$ and $F(1,2,6-9)=-0.063$ in REBO-2002E to fit the graphene edge energies at the first step of revision of the potential, the formation energy of chains becomes 0.96 eV, i.e. within the calculated range. Changing the parameter $F(2,3,9)$ to fit the barrier for vacancy migration in REBO-2002EV at the second step of revision of the potential does not affect this result. Therefore, no further manipulations are required to fit the REBO-2002 potential.

As for REBO-1990, adjusting only parameters $F(1,2,2)$ and $F(2,3,2)$ in REBO-1990EV to fit the graphene edge energies and barrier to vacancy migration at the first and second steps of revision of the potential, respectively, does not modify the formation energy of chains. To fit the latter in REBO-1990EVC we set the parameter $F(1,1,2) = 0.02514718$ and then readjust $F(1,2,2)$ and $F(2,3,2)$ again (see Table 1 and Table 2). The resulting formation energy of chains per atom is 0.95 eV.

To conclude, changing just several parameters in the REBO-1990 and REBO-2002 potentials [46,47] has allowed to improve significantly description of graphene edges, atomic carbon chains as well as migration of monovacancy in graphene. The corresponding versions REBO-1990EVC and REBO-2002EV perform better than other available potentials for carbon, such as ReaxFF$_{C-2013}$ [48] and LCBOPII [49,53]. REBO-1990EVC, nevertheless, is more accurate in the energies of graphene edges compared to REBO-2002EV. Furthermore, REBO-1990 and potentials derived from REBO-1990 have been extensively tested in simulations of transformations of carbon nanostructures [51,52,55-58] in recent years. Therefore, in the following we use REBO-1990EVC for studies of GNR evolution at high temperature.

*2.2 Details of calculations using interatomic potentials*

For calculations using interatomic potentials we use the in-house MD-kMC (Molecular Dynamics – kinetic Monte Carlo) code [95] in the case of REBO potentials (Refs. 46, 47 and 51 and the present paper) and LAMMPS [96] for ReaxFF$_{C-2013}$ (Ref. 48). First the geometry of an infinite graphene layer is optimized. To study the vacancy structure and migration the orthogonal simulation box with 10 unit cells along the armchair direction and 18 unit cells along the zigzag direction is considered under periodic boundary conditions. The size of the simulation box is optimized by the method of conjugated gradients till the energy change in consecutive iterations becomes less than $10^{-10}$ eV/atom. One of the atoms is removed from the simulation box and the initial guess for the structure corresponding to the 5/9 vacancy or the spiro state is provided. The positions of atoms within the unit cell are then optimized using the same method and stopping criterion.

To calculate the edge energies a 20-AGNR and a 36-ZGNR are constructed with 18 and 10 unit cells in the periodic direction, respectively, using the bond length in graphene obtained previously. To get the model of the reconstructed zigzag edge all hexagons at the edges of the 36-ZGNRs are converted into alternating pentagons and heptagons. The structures of the GNRs are geometrically optimized keeping the size of the simulation box fixed.

The energy of formation of atomic carbon chains is calculated for the simulation box with 10 atoms in the periodic direction. The atoms are initially displaced from the equidistant positions to make possible observation of the Peierls instability [97] and then optimization of atomic positions and the size of the simulation box in the periodic direction is performed.

MD simulations of transformation of GNR to chains have been carried out using REBO-1990EVC. Short 3-ZGNR and 4-ZGNR with the length of 88 Å consisting of 204 and 272 atoms, respectively, and a long 3-ZGNRs with the length of 260 Å consisting of 600 atoms have been considered. The atoms (12 atoms for 3-ZGNR and 16 atoms of 4-ZGNR) at both ends of the GNR are fixed. The velocity Verlet integration algorithm [98,99] with the time step of 0.6 fs is used. The temperature is maintained at $T$=2500 K by the Berendsen thermostat [100] with the relaxation time of 0.03 ps. To detect bond breaking and formation the topology of the carbon bond network is analyzed using the "shortest-path" algorithm [101] every 0.05 ps. Two atoms are considered as bonded if the distance between them is within 1.8 Å. The statistics on the time intervals between successive reactions and their localization is obtained.

*2.3 Details of DFT calculations*

To evaluate the formation energy of atomic carbon chains from graphene we have performed the spin-polarized DFT calculations using the PBE (Ref. 73) and PW91 (Ref. 74) exchange-correlation functionals as implemented in the VASP code [102]. The interaction of valence electrons with atomic cores is described by the projector augmented-wave method [103] in the case of the PBE functional and ultrasoft pseudopotentials [104] in the case of the PW91 functional. The maximal kinetic energy of the plane-wave basis set is 600 eV. A second-order Methfessel-Paxton smearing [105] with a width of 0.1 eV is applied. The integration over the Brillouin zone is carried out using the Monkhorst-Pack method [106]. The unit cells with periodic boundary conditions are considered. The vacuum gap between periodic images is 20 Å. For carbon chains, the unit cell comprises 2 atoms to include the Peierls instability [97] in consideration and 36 k-points are used along the chain. For graphene, the orthorhombic unit cell with 4 atoms is studied and the k-point grid with 24 k-points in the armchair direction and 36 k-points in the zigzag direction is used. The optimization of the unit cell size and structures within the unit cells is performed till the maximal residual force of 0.001 eV/Å.

To confirm the atomistic mechanism of GNR evolution that follows from MD simulations with the interatomic potential REBO-1990EVC the pathways of principal reactions leading to chain formation have been analyzed at the DFT level. The spin-polarized DFT calculations have been performed using the PBE functional [73]. The maximal kinetic energy of the plane-wave basis set is 400 eV. Integration over the Brillouin zone is carried out using the 3 × 1 × 1 k-point sampling. To check the influence k-point sampling we have also performed calculations with 3 different k-points meshes: 3 × 1 × 1, 6 × 1 × 1 and 9 × 1 × 1. For the periodic 3-ZGNR, the total energy of the system varies less than by 0.01 eV for these 3 samplings, so the chosen k-point mesh is sufficient for our calculations. The structures are geometrically optimized until the residual force acting on each atom becomes less than 0.03 eV/Å. The activation barriers are calculated with the nudged elastic band (NEB) method [107]. The unit cell of the periodic graphene ribbon is considered in the optimized 39.67 Å × 25 Å × 20 Å model cell (96 and 128 atoms in model cell for 3-ZGNR and 4-ZGNR, respectively), so that the spaces between the GNRs in neighbor model cells for aperiodic directions exceed 10 Å. The same initial structure and approach are employed to calculate the reaction barriers using the REBO-1990EVC potential.

## 3. Results and discussion

*3.1 MD simulations of chain formation*

To study the chain formation from ZGNRs under heat treatment we have performed reactive MD simulations using the revised REBO-1990EVC potential. Due to the influence of thermodynamic fluctuations, nonequilibrium transformation processes in nanoscale systems can give rise to different final nanoobjects for similar starting conditions [51,55,58]. Thus, at least several tens of simulation runs are necessary to obtain statistically significant results. In total, we have performed 50 MD simulation runs for the short 3-ZGNR and 4-ZGNR and 20 runs for the long 3-ZGNR to evaluate probabilities of formation of different nanoobjects.

*3.1.1. Final structures and transformation times for 3-ZGNR*

The most interesting finding of the MD simulations is that heating of 3-ZGNRs leads to formation of triple parallel chains attached to fixed atoms at both of the GNR ends for majority of the MD runs for the short 3-ZGNR (37 out of 50) and a significant part of the runs for the long 3-ZGNR (6 out of 20). An example of such structure evolution for the long 3-ZGNR is presented in Figure 1a. Figure 1a shows that formation of triple chains starts at several different places with subsequent merging of the growing chains. Thus, the time of 3-ZGNR transformation into triple chains (which corresponds to the moment when all the atoms of the system except the fixed ones become two-coordinated) does not depend on the GNR length if the latter is greater than several tens of atoms. The calculated average transformation times into triple chains are 29 ± 9 and 33 ± 10 ps, for the short and long 3-ZGNRs, respectively.

Formation of triple chains is the result of breaking of cross-bonds between atoms of neighbor zigzag atomic rows of 3-ZGNRs. Events of breaking of longitudinal bonds are much scarcer. They take place almost exclusively in pentagons, which are formed in reaction of cross-bond formation reverse to cross-bond breaking. When this happens, one or two parallel chains attached to the fixed atoms at both of the GNR ends are observed at the end of the simulation run (Figure 1b). Breaking of longitudinal bonds can in principle provoke a complete GNR rupture. Nevertheless, this was observed in only one run for the long 3-ZGNR (the structure evolution during this run is presented in Figure S1 in Supplementary data).

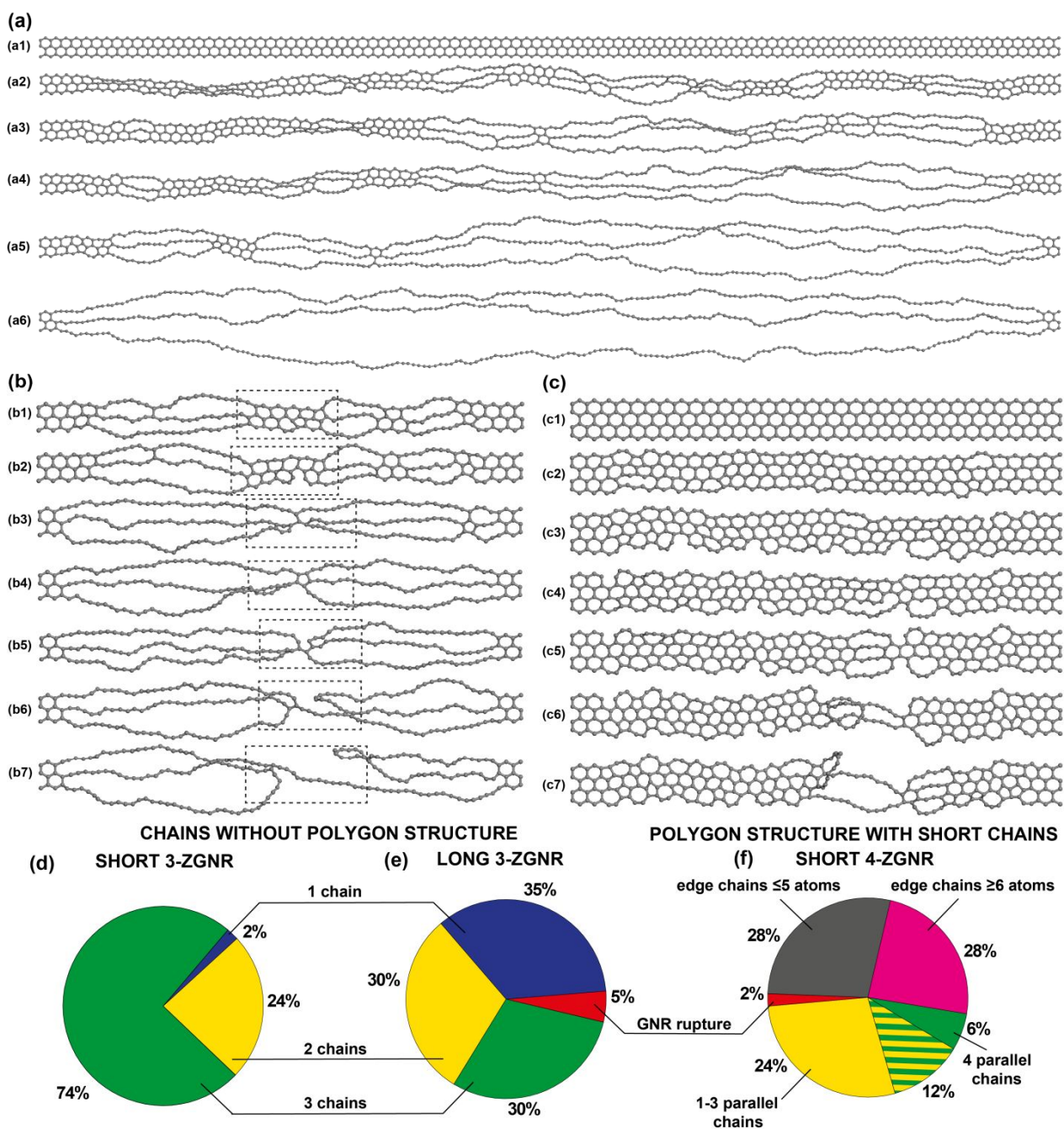

Figure 1: Simulated structure evolution of zigzag graphene nanoribbons with 3 atomic rows (3-ZGNR) and 4 atomic rows (4-ZGNR) under heat treatment at temperature 2500 K: (a) long 3-ZGNR, (a) short 3-ZGNR and (c) short 4-ZGNR. The structures shown for the long 3-ZGNR correspond to (a1) 0 ps, (a2) 6.0 ps, (a3) 8.92 ps, (a4) 12.65 ps, (a5) 21.67 ps, (a6) 43.74 ps. The structures for the short 3-ZGNR correspond to (b1) 7.4 ps, (b2) 7.86 ps, (b3) 17.5 ps, (b4) 21.16 ps, (b5) 24.34 ps, (b6) 24.89 ps and (b7) 25.19 ps. The regions of the structure inside the dotted frame in panel (b) correspond to scheme (a) of Figure S3 in Supplementary data. The structures for the short 4-ZGNR correspond to (c1) 0 ps, (c2) 10 ps, (c3) 35 ps, (c4) 75 ps, (c5) 80 ps, (c6) 85 ps and (c7) 100 ps. The distribution of the numbers of chains attached to the fixed atoms at both of the GNR after heat treatment of the long (d) and short (e) 3-ZGNR: 3 (green), 2 (blue) and 1 (yellow) chains and GNR rupture (red). (f) The distribution of the structures obtained after heat treatment of the short 4-ZGNR during 100 ps: structures containing pentagons, hexagons and heptagons with atomic chains only at the edge consisting of $\leq 5$ (grey) and $\geq 6$ (magenta) atoms, and with the regions consisting of 1-3 (yellow) and 4 (green) parallel chains and rupture of the GNR (red).

The distributions of the numbers of parallel chains attached to the fixed atoms at both of the GNR ends, which are formed as a result of the GNR heat treatment, are shown in Figure 1d and Figure 1e for the short and long 3-ZGNRs, respectively. The transformation time for the runs where the final structures contain single and double parallel chains or correspond to the complete GNR rupture is defined according to the moment when two fixed ends of the system are connected only through two-coordinated atoms or the link between them is lost. In these cases, the transformation time is determined by local structure rearrangement near broken longitudinal bonds and the average transformation times reach $79 \pm 23$ and $133 \pm 25$ ps, for short and long 3-ZGNR, respectively. Since the transformation time for the single and double chain formation is considerably greater than for the triple chain formation, it is clear that the local structure evolution near broken longitudinal bonds is much slower than propagation of triple chains. The atomistic mechanisms of the origin and growth of triple chains are considered in detail below. The merging of triple chains and formation of structural fragments dangling at only one of the GNR ends are described in Supplementary data.

*3.2.2. Atomistic mechanism of formation of triple parallel chains*
Let us now consider the atomistic mechanism of formation of triple parallel chains from the 3-ZGNR. For this purpose, the total numbers of the reactions of cross-bond breaking and reverse reactions of cross-bond formation have been calculated for all 37 runs where triple chain formation from the short 3-ZGNR is observed (Figure 2). The calculated average number of reactions before the GNR is completely decomposed into three chains is about $82 \pm 10$, which corresponds to about 70 and 10 reactions of cross-bond breaking and formation, respectively. The reverse reaction of cross-bond formation commonly leads to formation of a pentagon or a pentagon-heptagon pair for the shortest chain length of 3 atoms. However, the presence of polygons different from hexagons does not contribute noticeably into the statistics of cross-bond breaking. For example, out of 423 and 208 reactions of the first and second cross-bond breaking, only 19 and 17, respectively, are not between two hexagons. Such rare reactions are not counted in Figure 2a. Any reactions which lead to merging of neighbor chains are not included either. Because of this, the net numbers of incoming and outgoing reactions for structures shown in Figure 2a are not the same.

The dominant pathway of triple chain formation is shown in Figure 2 by green arrows and corresponds to the following sequence of structures: [0,0], [1,0], [1,1], [2,1], [2,2], and so on, where [$n,m$] denotes the structure with and $n$ and $m$ broken cross-bonds in the upper and lower rows of hexagons. Along this dominant pathway, cross-bond breaking in one of the rows of hexagons is followed by cross-bond breaking in the second row. The reactions where the cross-bond is broken in the same row of hexagons as at the previous step are more than an order of magnitude less frequent (the secondary pathway shown in Figure 2 by blue arrows). The dominant pathway of formation of triple chains revealed by MD simulations is confirmed below by DFT calculations of activation barriers for the reactions of cross-bond breaking.

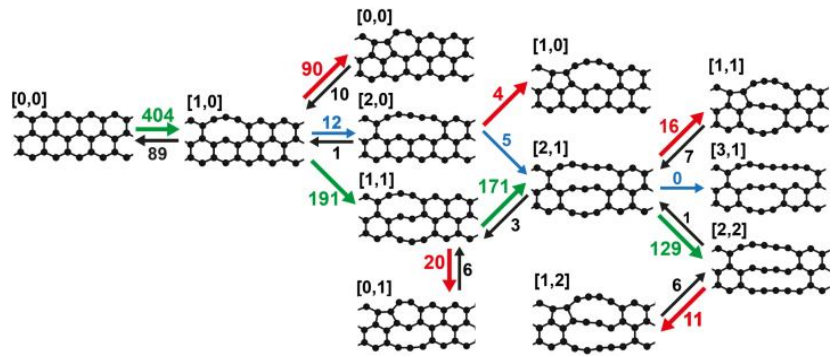

Figure 2: Scheme of origin and initial stage of formation of triple parallel chains under heat treatment of 3-ZGNRs. The numbers of broken cross-bonds in the upper and lower rows of hexagons for the structures at the initial stage of chain formation are indicated in square brackets. The calculated numbers of reactions of cross-bond formation and breaking corresponding to transitions between the structures given are indicated. With account of polygons different from the ones shown in the picture, the total number of cross-bond breaking events is 423 at the first step [0,0] →[1,0] and 208 at the second step [1,0] →[1,1]. The reactions of cross-bond breakings corresponding to the dominant and secondary pathways of triple chain formation are shown by green and blue arrows, respectively. The reactions of cross-bond formation leading to generation of pentagons are shown by red arrows.

*3.2.3. Formation of single and double chains as a result of longitudinal bond breaking*

As noted above, formation of pentagons in reactions of cross-bond formation can be essential for breaking of longitudinal bonds between atoms of the same zigzag row. While after the first cross-bond breaking, about half of the reverse reactions restore two hexagons, later the fraction of hexagons formed in the reverse reactions is only about 15%. Thus, reactions of cross-bond formation efficiently generate pentagons. Before the third cross-bond breaking, formation of pentagons takes place in the form of pentagon-heptagon pairs. In total, however, heptagons accompany pentagons only in half of the reverse reactions. Only in 10 out of 90 reactions of pentagon-heptagon pair formation after the first cross-bond breaking, this bond is broken again. This is similar to the probability for cross-bond breaking between two hexagons and it can be concluded that pentagon-heptagon pairs and hexagon pairs in the 3-ZGNR have comparable stabilities relative to cross-bond breaking at high temperature. The reactions of cross-bond formation are observed up to structure with one cross-bond remaining in each row. Once one of these cross-bonds is broken, the second one is also broken irreversibly. This is consistent with the result of DFT calculations that the 2-ZGNR has a considerably greater formation energy than two atomic chains [22]. The scheme of triple chain merging and analysis of statistics of reactions of cross-bond breaking and formation at triple chain merging is given in Supplementary data.

Let us now consider simulation runs where less than three atomic chains attached to fixed atoms at the both GNR ends are formed. In all such runs, there are local segments of 3-ZGNR where at least one longitudinal bond between atoms of the same zigzag row is broken during structure rearrangement without bond recovery. One and two of such local segments have been found in 8 and 5 simulation runs, respectively, out of 50 runs in total for the short 3-ZGNR with 60 atoms in the zigzag atomic row. This means that there should be on average one longitudinal bond breaking per 170 atoms of the zigzag atomic row. Given this probability $p = 1/170$, the probability $P = 1/2$ that a triple chain is formed in half cases is achieved for the GNR with $N_{1/2} = \ln(1/2)/\ln(1-p) \approx 1/(p \ln 2) \approx 120$ atoms in the zigzag atomic row. Triple chains have been obtained in 6 simulation runs out of 20 for the long 3-ZGNR with 200 atoms in the zigzag atomic row, which is in excellent agreement with this rough estimate. The schemes with examples of structure evolution after the first longitudinal bond breaking are shown in Figure S3 in Supplementary data. In the majority of the cases, the first longitudinal bond breaking happens in the external zigzag atomic row and the broken bond belongs to a pentagon. The first longitudinal bond breaking in the internal zigzag atomic row is found only in one simulation run. Using the statistics on breaking of longitudinal bonds we can estimate that the maximal length of single chains can reach 1000 atoms (see Supplementary data).

If at least one longitudinal bond is broken without recovery, not only anchored chains attached to fixed atoms at the both ends of the 3-ZGNR but also dangling chains attached to fixed atoms only at one end of 3-ZGNR (Figure 1b) as well as occasionally free chains are formed. The detailed description of these dangling and free chains observed in all simulation runs for the short and long 3-ZGNRs is presented in Table S1 in Supplementary data. The formation of chains with a free end always occurs via breaking of a bond at the chain end. This result of our simulation is in agreement with experimental observations [18,22,26,27]. The detachment of atomic carbon chain from fullerene by breaking of the end bond has been also observed in tight-binding MD simulations [18].

*3.1.4. Results for 4-ZGNR*

To compare the structural transformation of 3-ZGNRs and 4-ZGNRs, heating of the short 4-ZGNR at the same temperature has been simulated in MD runs of 100 ps duration, which exceeds the average transformation time for 3-ZGNRs. Surprisingly, it turned out that just one additional zigzag row of atoms leads to the essentially different structure evolution (see an example in Figure 1c). Different from the case of 3-ZGNRs, only short chains at the 4-ZGNR edges arise within the same time of 30–70 ps (Figure 1c3 and Figure 1c4). The efficient generation of pentagon-heptagon pairs and breaking of longitudinal bonds at the 4-ZGNR edges (that is GNR etching) take place simultaneously with the formation of short chains. The GNR etching prevents formation of tetrad chains in 4-ZGNR. Only short regions with tetrad chains (from 2 to 14 atoms in length) are found during further heating of the 4-ZGNR in 5 runs out 50 (see, for example, structures presented in Figure 1c5 and Figure 1c7). The distribution of final structures of the 4-ZGNR after the heat treatment during 100 ps is shown in Figure 1f.

The chain formation in relatively wide GNRs has been observed in experiments under electron irradiation [22,23,25] and combined influence of electron irradiation and Joule heat [24] and studied in previous MD simulations for heat treatment [35] and electron irradiation [37]. The MD simulations for the 4-ZGNR performed here confirm the following qualitative characteristics of the process revealed previously: formation of pentagon-heptagon pairs starting from the GNR edge [22,25,35,37], etching of the GNR with decrease of its width [22-25], formation of multiple parallel chains with length up to several tens atoms [22,23,25,35]. Note that listed above qualitative features of chain formation are the same for HRTEM studies [22-25] and MD study here. The similarity of transformation processes under heat treatment and electron irradiation was observed previously in experimental studies of GNR formation inside carbon nanotubes [108] and MD simulations of the graphene-fullerene transformation [55,109]. The relation between the rates of different processes in our simulations is consistent with the experimental observation of etching of a GNR formed between two holes in a graphene layer under electron irradiation [25]. Namely, it was observed that etching of edges of the GNR is a slow process while its width exceeds 4 atomic rows. However, as soon as some part of the GNR reaches the width of 3 atomic rows, the decomposition of this part into chains is very fast.

It is clear that the structural transformation of both 3-ZGNR and 4-ZGNR, while leads to very different final structures, is determined by the interplay of the same processes of breaking of cross-bonds and longitudinal bonds. The reasons of the very distinct structural transformation should be looked for in the difference of the corresponding rates and they are considered in Section 3.2 based on DFT calculations of the activation barriers of cross-bond breaking and formation of pentagon-heptagon pairs.

*3.2 Analysis of reaction pathways*

To confirm the atomistic mechanism of triple chain formation during the heat treatment of the 3-ZGNR revealed in the MD simulations we have analyzed the potential energy surfaces for principal reactions leading to chain formation. Namely, the reactions at the initial stage of triple chain formation from the 3-ZGNRs, of single chain formation at the edge of the 4-ZGNR and of pentagon-heptagon pair formation at the edges of both of these GNRs are considered (Figure 3 and Tables 3 – 5). In addition to the barrier, $E_a$, we also consider the energy change, $\Delta E_1$, and the energy of the product relative to the pristine GNR, $\Delta E_2$, i.e. the formation energy of the product.

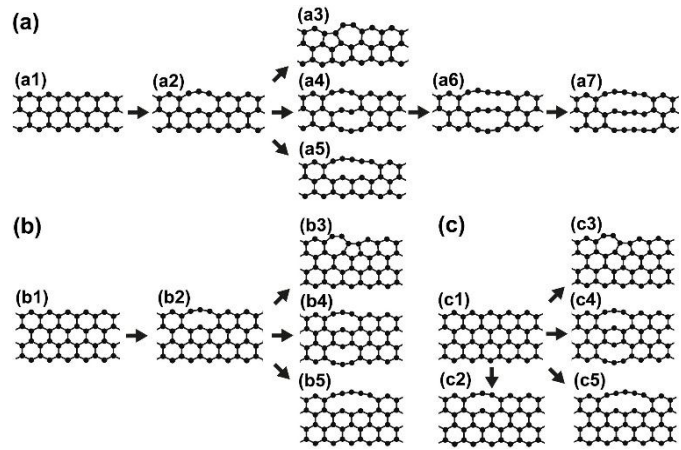

Figure 3. Schemes of principal reactions at the initial stage of chain formation from the 3-ZGNR (a) and (b),(c) 4-ZGNR. Calculated barriers for the reactions presented in panels (a), (b) and (c) are given in Tables 3, 4 and 5, respectively. Directions of reactions are indicated by arrows.

First let us compare the energetic characteristics of local structures at the edge of narrow ZGNRs obtained here by the DFT calculations with similar data from literature. While formation of pentagon-heptagon pairs at the zigzag graphene edge has been actively investigated in recent years using *ab initio* methods [25,53,60,61,62,63], we are not aware of such studies for formation of chains at graphene edges. The barriers and energy changes for simultaneous and complete reconstruction of the zigzag edge [53,60,61] as well as formation of the first [25,62,63] and second [62] pentagon-heptagon pairs have been reported. The barriers of 1.12 eV and 1.80 eV were obtained for formation of the first pentagon-heptagon pair at the edge of the 4-ZGNR in Refs. 25 and 62, respectively. Such a difference in the results of the same authors can be explained by insufficiently large supercells used. In Ref. 25, the supercell included only 4 hexagons along the GNR edge. In Ref. 62, the formation of the first pentagon-heptagon pair at the free edge was found to lead to the energy increase of 0.2 eV, while the formation of the second pair next to the previous one to the energy reduction of 0.7 eV. The barrier of 1.11 eV was predicted for the second step. In calculations [63] for the 7-ZGNR with 8 hexagons along the edge, the formation of the first pentagon-heptagon pair was found to be energetically favourable with the energy release of 0.21 eV and the barrier of 1.61 eV. The energy released was found to change from 0.18 eV to 0.24 eV upon varying the ZGNR width from 6 to 12 zigzag rows. Such a behaviour was explained by a better relaxation of the stress induced by formation of the pentagon-heptagon pair for wider GNRs.

We have obtained close results for the barrier for the formation of the first pentagon-heptagon pair in narrow ZGNRs. The corresponding values are 1.92 and 1.65 eV for the 4-ZGNR (reaction (c1)-(c3) in Figure 3, Table 5) and the 3-ZGNR (reaction (a1)-(a2)-(a3) in Figure 3, Table 3), respectively. The energy release upon the formation of the first pentagon-heptagon pair is somewhat greater than in Ref. 63 and reaches 0.32 and 0.69 eV for the 4-ZGNR (Table 5) and 3-ZGNR (Table 3), respectively. Therefore, according to our calculations, generation of the first pentagon-heptagon pair occurs more favourably, with a smaller barrier and a greater energy released, in the narrower 3-ZGNR compared to the 4-ZGNR.

Table 3. Energy characteristics for the reactions at the initial stage of chain formation from the 3-ZGNR obtained using the interatomic potential and by the DFT calculations: barriers $E_a$, energy change $\Delta E_1$ and energy $\Delta E_2$ of the product relative to the pristine 3-ZGNR (in eV). The notation [n,m] denotes the structure with and n and m broken cross-bonds in the upper and lower rows of hexagons, [0,0]5/7 denotes the structure with a single pentagon-heptagon pair. The structures shown in Figure 3a are also indicated.

| | | DFT (PAW PBE) | | | Potential REBO-1990EVC | | |
|---|---|---|---|---|---|---|---|
| Reaction | Figure 3a | $E_a$ | $\Delta E_1$ | $\Delta E_2$ | $E_a$ | $\Delta E_1$ | $\Delta E_2$ |
| [0,0]-[1,0] | (a1)-(a2) | 1.64 | 1.49 | 1.49 | 1.75 | 1.38 | 1.38 |
| [1,0]-[0,0]5/7 | (a2)-(a3) | 0.01 | −2.18 | −0.69 | 0.60 | −1.68 | −0.29 |
| [1,0]-[1,1] | (a2)-(a4) | 0.14 | −1.34 | 0.15 | 1.06 | −0.40 | 0.98 |
| [1,0]-[2,0] | (a2)-(a5) | | unstable | | 2.00 | 1.81 | 3.20 |

Let us now compare the energetic characteristics of the reactions obtained by the DFT calculations and using the potential (Tables 3 – 5). The potential describes the correct order of energies of the considered structures both for the 3-ZGNR and 4-ZGNR including the relative stability of similar structures for the GNRs of different width. Although the stable states of structures (a5) and (b2)/(c2) are found only using the potential, this does not contradict the revealed atomistic mechanism of the triple parallel chain formation. The difference in the values obtained by the DFT calculations and using the potential is within 0.5 eV for the majority of the calculated energy characteristics. Although in several cases, this difference exceeds 0.5 eV, as discussed below, the values obtained by the DFT calculations should be even more favourable for triple chain formation from 3-ZGNR and should make the distinction between the evolution of the 3-ZGNR and 4-ZGNR even more prominent in comparison with the interatomic potential.

Table 4. Energy characteristics for the reactions at the initial stage of chain formation from the 4-ZGNR calculated using the interatomic potential REBO-1990EVC: barriers $E_a$, energy change $\Delta E_1$ and energy $\Delta E_2$ of the product relative to the pristine 4-ZGNR (in eV). The notation [n,m] denotes the structure with and n and m broken cross-bonds in the upper and lower rows of hexagons, [0,0]5/7 denotes the structure with a single pentagon-heptagon pair. The structures shown in Figure 3b are also indicated.

| Reaction | Figure 3b | $E_a$ | $\Delta E_1$ | $\Delta E_2$ |
|---|---|---|---|---|
| [0,0]-[1,0] | (b1)-(b2) | 1.84 | 1.39 | 1.39 |
| [1,0]-[0,0]5/7 | (b2)-(b3) | 0.62 | −1.52 | −0.13 |
| [1,0]-[1,1] | (b2)-(b4) | 1.63 | 1.30 | 2.70 |
| [1,0]-[2,0] | (b2)-(b5) | 2.00 | 1.78 | 3.20 |

Table 5. Energy characteristics for the reactions at the initial stage of chain formation from 4-ZGNR obtained by the DFT calculations: barriers $E_a$, energy change $\Delta E_1$ and energy $\Delta E_2$ of the product relative to the pristine 4-ZGNR (in eV). The notation [n,m] denotes the structure with and n and m broken cross-bonds in the upper and lower rows of hexagons, [0,0]5/7 denotes the structure with a single pentagon-heptagon pair. The structures shown in Figure 3c are also indicated.

| Reaction | Figure 3c | $E_a$ | $\Delta E_2$ |
|---|---|---|---|
| [0,0]-[1,0] | (c1)-(c2) | unstable | |
| [1,0]-[0,0]5/7 | (c1)-(c3) | 1.92 | −0.32 |
| [0,0]-[1,1] | (c1)-(c4) | 3.90 | 3.76 |
| [0,0]-[2,0] | (c1)-(c5) | 4.90 | 4.88 |

The new version of the Brenner potential REBO-1990EVC is more accurate in description of formation of chains and pentagon-heptagon pairs at the graphene edge compared to other available interatomic potentials. While the barrier and energy change for formation of a three-atom chain at the 3-ZGNR edge agree with the DFT data within just 0.1 eV (Table 3), another version of the Brenner potential, AIREBO [110], gives the values of 3.1 and 2.6 eV, respectively [64], which are almost twice greater than our DFT results. The original REBO-1990 (Ref. 46) also gives the barriers on the order of 3 eV for various reactions that require breaking of only one bond, including the formation of chains and pentagon-heptagon pairs at the zigzag graphene edge [56,57]. The latter potential erroneously predicts that formation of a pentagon-heptagon pair is unfavourable by 1 eV. The modified LCBOPII potential [53] gives the free-energy barrier of only 0.83 eV for formation of the first pentagon-heptagon pair at room temperature, which is in line with the DFT values [53,60,61] for the simultaneous edge reconstruction but far below the DFT results for the first pair (see Refs. 62, 63 and Table 3, Table 5).

The energy characteristics of principal reactions during chain formation obtained by the DFT calculations are consistent with the atomistic mechanism of the triple parallel chain formation under heat treatment of 3-ZGNR established in the previous section by the MD simulations. First let us consider the energetics of the 3-ZGNR with two broken cross-bonds (structure (a4) in Figure 3a), which is a nucleus for further propagation of triple chains according to the revealed atomistic mechanism. Both calculations based on DFT and the potential show that 1) structure (a4) is energetically more favorable than structure (a2) with one broken cross-bond and 2) structure (a4) has a large barrier for the reverse reaction of cross-bond formation (1.48 eV and 1.46 eV according to the DFT calculations and using the potential, respectively). Moreover, the energy of structure (a4) relative the pristine 3-ZGNR (a1) and relative to structure (a3) with a single pentagon-heptagon pair obtained by the DFT calculations is even smaller than the one obtained using the potential. Thus, formation of structure (a4) at high temperature is even more probable according to the DFT calculations than in the model based on the REBO-1990EVC potential.

To address the energetics of reactions in the 3-ZGNR at the stage of triple parallel chain propagation the energetics of the dominant pathway up to the fourth cross-bond breaking has been analyzed using the REBO-1990EVC potential. Figure 4 shows that both the barriers of 1.14 and 1.15 eV and the energy release of 0.13 and 0.11 eV at breaking of the 3-rd and 4-th cross-bonds, respectively, are nearly equal. Therefore, it can be expected that further cross-bond breaking events should have the same energy characteristics. It can be also noted that the barriers for these steps are close to the formation energy of chains from graphene of about 0.95 eV (see Section 2.1).

We can also compare the energetics of similar structures (a4) and (c4) (see Figure 3) of the 3-ZGNR and 4-ZGNR, respectively, with two broken cross-bonds (which can be considered as nuclei for further growth of chains). The energy cost of structure (c4) relative the pristine 4-ZGNR is much greater than the energy cost of structure (a4) relative the pristine 3-ZGNR, 3.76 eV versus 0.15 eV by the DFT calculations and 2.7 eV versus 0.98 eV for the REBO-1990EVC potential. Simultaneously, the barrier for the reverse reaction of cross-bond formation for structure (c4) of the 4-ZGNR is considerably smaller than for structure (a4) of the 3-ZGNR, 0.15 eV versus 1.46 eV according to the DFT calculations and 0.33 eV versus 1.48 eV for the potential. Thus, the probability of formation and stability at high temperature for structure (c4) of the 4-ZGNR is much smaller than for similar structure (a4) of the 3-ZGNR. In the both cases, the DFT calculations predict even a slightly more pronounced difference compared to the interatomic potential. On the other hand, both of the approaches agree that formation of long chains by consecutive cross-bond breaking in only one hexagon row is highly unfavorable (a5,b5,c5) and the difference in the parameters describing generation of pentagon-heptagon pairs (a3,b3,c3) in the 3-ZGNR and 4-ZGNR is small. Therefore, the DFT calculations not only confirm the atomistic mechanism of triple chain formation during heating of the 3-ZGNR but also explain the qualitatively distinct evolution of the 3-ZGNR and 4-ZGNR under heat treatment.

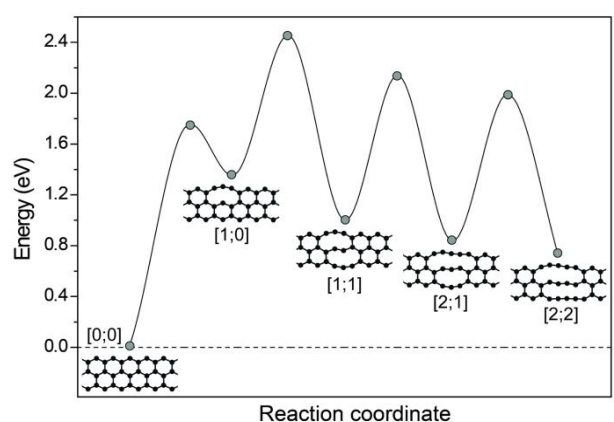

Figure 4: Schematic representation of energy variations along the dominant pathway at the initial stage of the triple parallel chain formation under heating of the 3-ZGNR calculated using the interatomic potential REBO-1990EVC. The numbers of broken cross-bonds in the upper and lower rows of hexagons for these structures are indicated in square brackets.

*3.3. Chain-based nanoelectronic device*

Single-molecule electronics is a fast developing field leading to advances in fundamental physics and chemistry and giving rise to various applications (see, for example, reviews on transport properties of molecular wires [111] and on single-molecule conductance switches [112]). Methods of production of single-molecule electronic nanodevices placing a molecular wire synthesized ex situ between electrodes and by bottom-up growth of molecular wires between electrodes have been elaborated (see Ref. [111] for a review). Here we propose that thermal treatment of GNRs can be used for top-down synthesis of molecular wires between electrodes. A method for GNR transfer from the substrate where the GNR is synthesized to a nanodevice so that the GNR bridges the gap between the electrodes has been implemented recently [34]. As shown in Section 3.1, heating of the GNR can lead to formation of anchored and dangling chains (Joule heat can be used). Such a system with both anchored and dangling chains can be beneficial for development of new single-molecule electronic devices. Since dangling chains are flexible and have a radical at their end, they can easily stick to any nanoobject brought in a close contact. For example, a possible scheme of a three-electrode nanodevice based on atomic carbon chains is shown in Figure 5. Initially the 3-ZGNR (4) is attached to two electrodes (1) and (2) and the third electrode (3) is located in the vicinity. If some of the dangling chains attach the third electrode (3), the three-electrode nanodevice is produced.

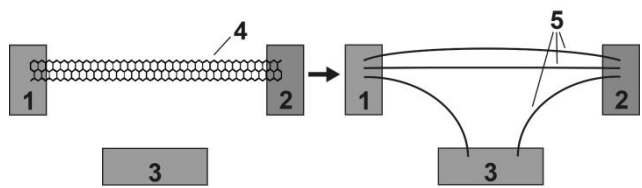

Figure 5. Schematic representation of the nanodevice based on atomic carbon chains (5) with three electrodes (1), (2) and (3), which can be fabricated by heat treatment of a 3-ZGNR (4) placed between electrodes (1) and (2).

## 4. Conclusions

The MD simulations carried out in the present paper demonstrate formation of long single, double and triple atomic carbon chains by heating of 3-ZGNRs at 2500 K. Based on the simulations performed, we predict that triple chains with the length up to several hundreds atoms and single chains with the length up to a thousand of atoms can be produced using this treatment. It should be noted, however, that only short chains of about 10 atoms in length can be obtained under heat treatment of 4-ZGNRs under the same conditions.

To perform the simulations several parameters of the Brenner potential of the first generation [46] have been readjusted to make it more adequate for modelling of structural transformations in graphene. The second-generation Brenner potential [47] has been also reconsidered. The updated versions of these potentials provide a much more accurate description of graphene edge energies, vacancy migration and formation energy of atomic chains compared to other available interatomic potentials for carbon.

The analysis of detailed statistics of the bond breaking and formation observed in MD simulations has been used to determine the atomistic mechanism of the single, double and triple chain formation at heat treatment of 3-ZGNR. The dominant pathway of triple chain formation is revealed. Along this pathway, breaking of cross-bond between atoms of the external and inner zigzag atomic rows in one of the rows of hexagons is followed by cross-bond breaking in the second row. The reactions where the cross-bond is broken in the same row of hexagons as at the previous step are at least an order of magnitude less frequent. The ratio of frequencies of the reactions of cross-bond breaking and formation is equal to approximately 7 and pentagons are predominantly formed in the reverse reactions of cross-bond formation. The revealed atomistic mechanism of triple chain formation is confirmed by the DFT calculations of the barriers of principal reactions leading to chain formation.

Formation of single or double chains is also possible during the heat treatment of 3-ZGNR and occurs as a consequence of breaking of at least one longitudinal bond between atoms within the same zigzag atomic row. In majority of the cases, the first event of longitudinal bond breaking takes place in the external zigzag atomic row and the broken bond belongs to a pentagon. Once the first longitudinal bond is broken, diverse scenarios of local structure rearrangement in the vicinity of this bond, mainly with subsequent breaking of additional longitudinal bonds, are observed. These local structural rearrangements normally do not lead to the complete rupture of the 3-ZGNR. Nevertheless, the presence of several segments with a broken longitudinal bond in a sufficiently long 3-ZGNR can provoke a GNR rupture. To study this process the 3-ZGNR length and number of simulation runs should be at least several times greater and this is beyond the scope of the present paper.

We believe that the proposed method of synthesis of atomic carbon chains by heat treatment of the 3-ZGNR has a promise for fabrication of chain-based electronic nanodevices. For example, Joule heating of 3-ZGNR with the ends attached to two electrodes can be used. According to the MD simulations performed, dangling chains attached to only one electrode can be produced under heat treatment along with anchored chains attached to both electrodes. If the third electrode is placed in the vicinity of the GNR heated, these dangling chains can stick by the free end with the radical atom to the third electrode. Thus, not only two-electrode but also three-electrode carbon chain-based electronic nanodevices can be obtained.

The method of generation of atomic chains from GNRs proposed in the present paper relies on the possibility of production of narrow GNRs with a homogeneous edge and width. Synthesis of such 6-ZGNRs [33], armchair GNRs with 7 atomic rows [28-30] or chiral (3,1)-GNRs [31,32] has been demonstrated recently using chemical methods. Though zigzag GNRs with 3 and 4 atomic rows studied in the present paper have not yet been obtained, there is no fundamental obstacle impeding synthesis of these structures in the nearest future.

**Acknowledgements**

ASS, AMP and AAK acknowledge the Russian Foundation of Basic Research (Grant 18-02-00985). IVL acknowledges Grupos Consolidados del Gobierno Vasco (IT-578-13) and EU-H2020 project "MOSTOPHOS" (n. 646259). This work has been carried out using computing resources of the federal collective usage center Complex for Simulation and Data Processing for Mega-science Facilities at NRC "Kurchatov Institute", http://ckp.nrcki.ru/.

# Supplementary data for
# "Long triple carbon chains formation by heat treatment of graphene nanoribbon: MD study with revised Brenner potential"

*Alexander S. Sinitsa[1], Irina V. Lebedeva[2*], Andrey M. Popov[3*], Andrey A. Knizhnik[4]*

[1] *National Research Centre "Kurchatov Institute", Kurchatov Square 1, Moscow 123182, Russia*

[2] *Nano-Bio Spectroscopy Group and ETSF, Universidad del País Vasco, CFM CSIC-UPV/EHU, San Sebastian 20018, Spain*

[3] *Institute for Spectroscopy of Russian Academy of Sciences, Fizicheskaya Street 5, Troitsk, Moscow 108840, Russia*

[4] *Kintech Lab Ltd., 3rd Khoroshevskaya Street 12, Moscow 123298, Russia*

**Corresponding Author**

E-mail: popov-isan@mail.ru

**Contents**

**Structure evolution in a single simulation run where the 3-ZGNR rupture takes place**

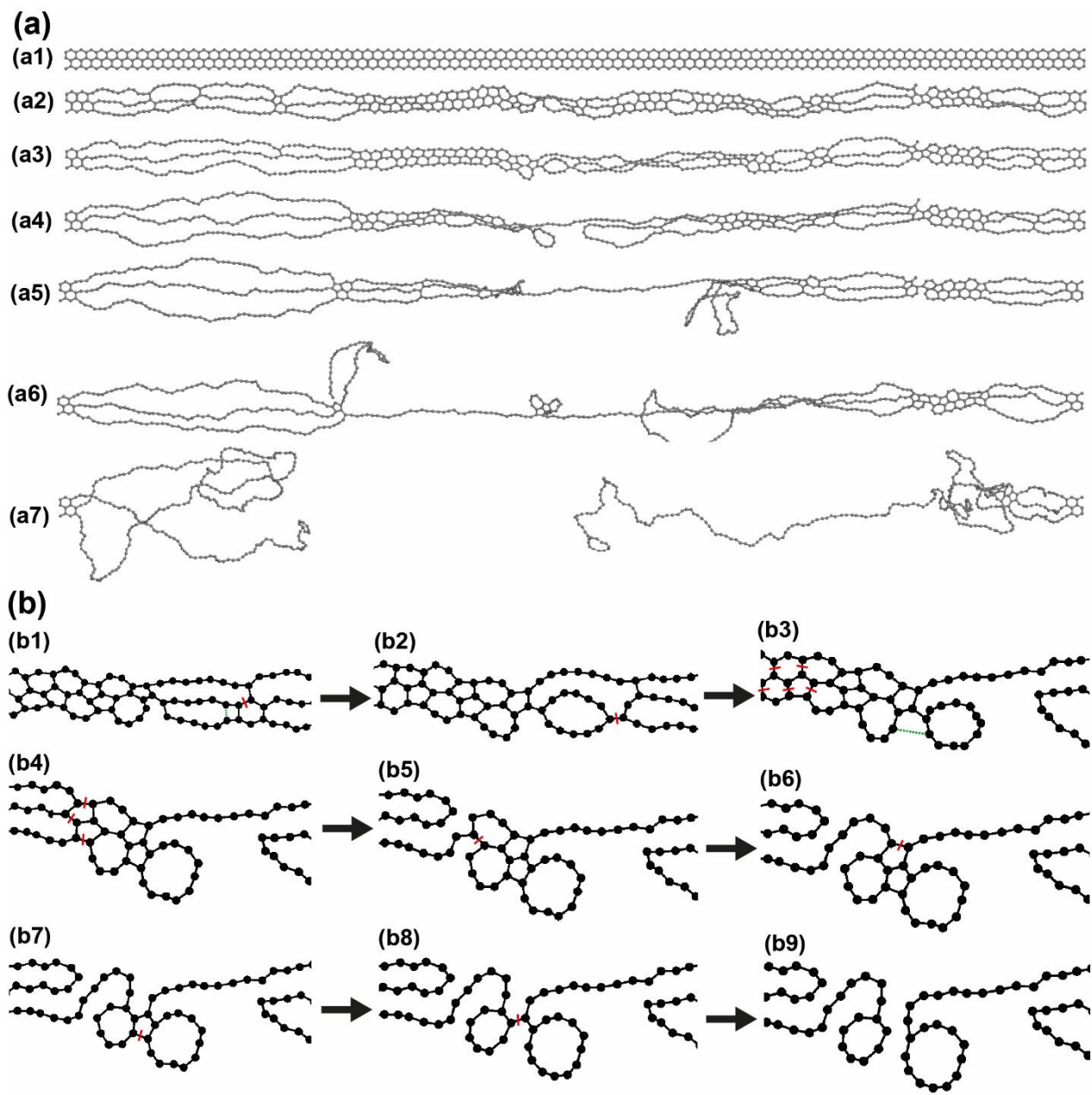

**Figure S1**: Simulated structure evolution of the zigzag-edged graphene nanoribbon with 3 atomic rows (3-ZGNR) under heat treatment at temperature 2500 K in a single simulation run with the 3-ZGNR rupture: (a) structures observed at (a1) 0 ps, (a2) 10.78 ps, (a3) 14.16 ps, (a4) 17.99 ps, (a5) 20.81 ps, (a6) 24.14 ps and (a7) 38.55 ps; (b) Schematic representation of the local structure evolution for the 3-ZGNR rupture. The broken bonds are crossed by red strokes. The formed bonds are shown by thin green lines.

**Scheme and analysis of reactions of cross-bond breaking and formation at the final stage (chain merging) of formation of triple parallel chains under heat treatment of 3-ZGNRs.**

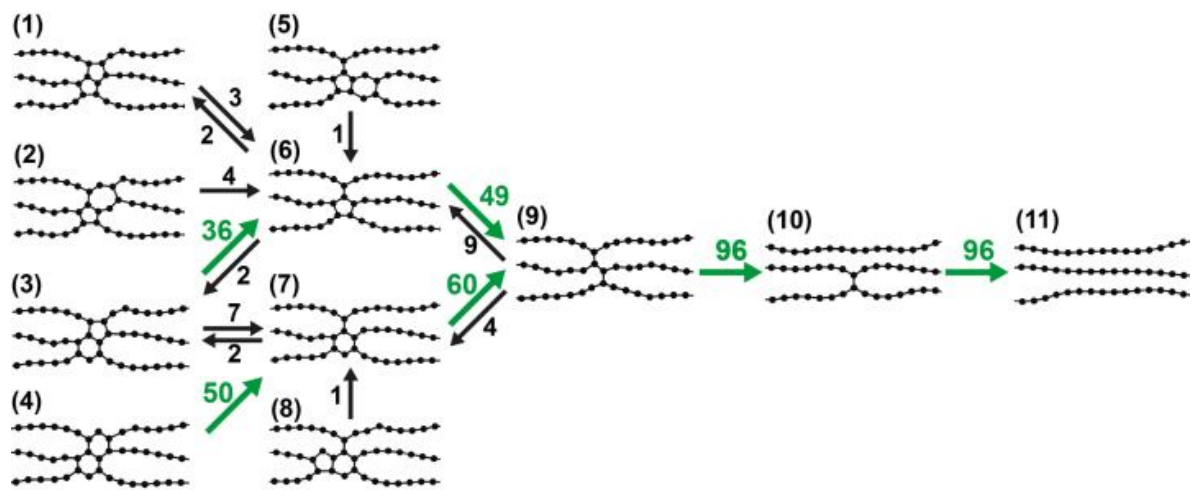

**Figure S2**: Scheme of the final stage (chain merging) of formation of triple parallel chains under heat treatment of 3-ZGNRs. The calculated numbers of reactions of cross-bond formation and breaking corresponding to transitions between the structures given are indicated. The reactions of cross-bond breaking corresponding to the dominant pathway of triple chain formation are shown by green arrows.

The total of 96 cases of triple chain merging observed in 37 runs where triple chains are formed corresponds to the average triple chain length of 23 atoms just before the merging. Figure 2b shows the scheme and statistics for the reactions of cross-bond breaking and formation for merging of triple chains of more than 5 atoms in length starting from the structures where just a pair of polygons is left between the merging triple chains. A quarter of these polygons are pentagons and only few hundredth part are heptagons. The portion of pentagons increases to about a half in structures with a single polygon left. Thus, it is possible to say that pentagons are slightly more stable than hexagons relative to cross-bond breaking. For example, the average lifetime of a single pentagon remaining is 2.9 ± 0.4 ps whereas the average lifetime of a single hexagon remaining is 1.2 ± 0.2 ps. However, such a difference in the lifetimes does not affect the qualitative features of the atomistic mechanism of the triple chain formation from the 3-ZGNR.

**Schematic representation of examples of the local structure evolution of 3-ZGNRs under heat treatment in the cases of formation of single and double parallel chains attached to fixed atoms at the both GNR ends**

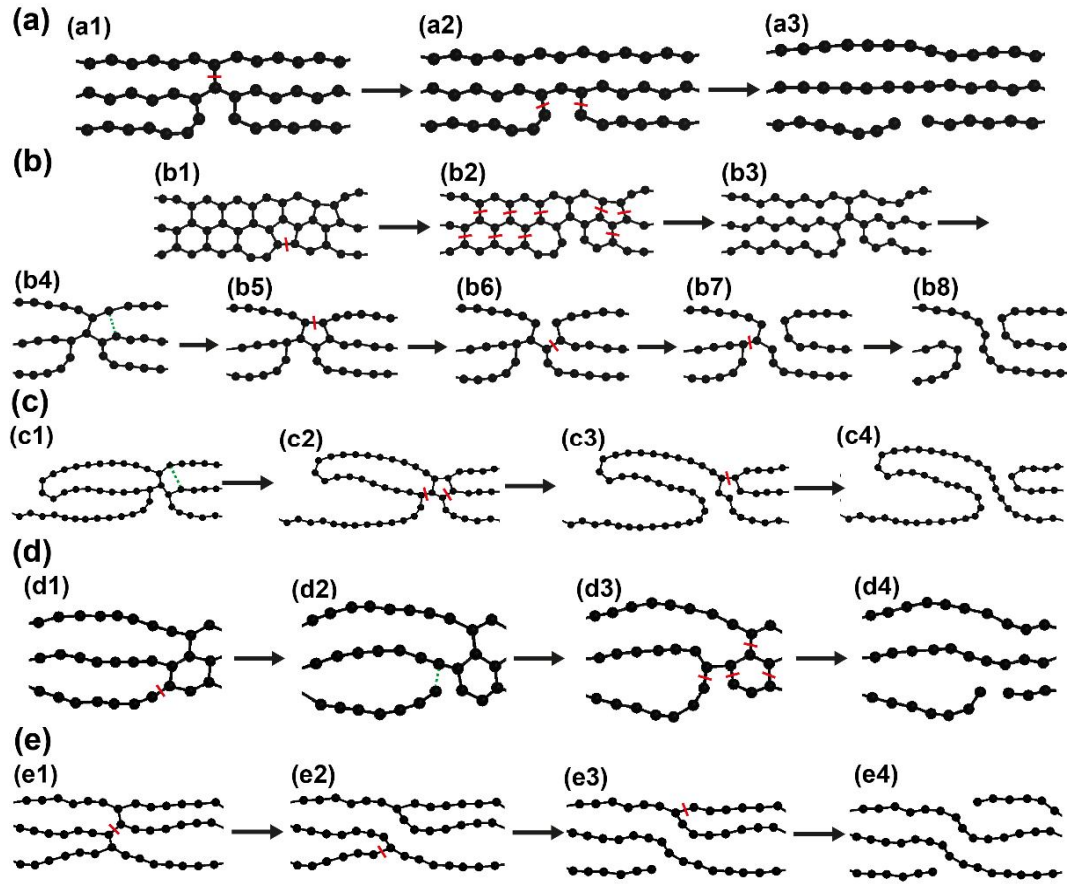

**Figure S3**: Schematic representation of examples of the local structure evolution of 3-ZGNRs under heat treatment at temperature 2500 K for the cases of formation of single and double parallel chains attached to fixed atoms at the both GNR ends. The broken bonds are crossed by red strokes. The forming bonds are shown by thin green lines.

The schemes with examples of structure evolution after longitudinal bond breaking are shown in Figure S3. Three different locations of the first event of longitudinal bond breaking are observed. In the majority of the cases, it happens in the external zigzag atomic row and the broken bond belongs to a pentagon (Figure S3a,b,c). The cases of local structure evolution with a single broken longitudinal bond which belongs to a pentagon and two broken longitudinal bonds which belong to adjacent pentagons from the upper and lower rows of polygons are shown in Figure S3b and Figure S3a, respectively. Longitudinal bond breaking at the end of an external chain followed by attachment of this chain to the parallel internal chain is observed in several simulation runs (see Figure 3d). It is interesting to note that in the case of two longitudinal bonds broken in different local segments of 3-ZGNR, formation of a single chain attached to fixed atoms at both ends of 3-ZGNR with a length greater than the GNR length is sometimes possible (Figure S3c). Longitudinal bond breaking at the end of an external chain followed by attachment of this chain to the parallel internal chain is observed in several simulation runs (Figure S3d). The first longitudinal bond breaking in the internal zigzag atomic row is found only in one simulation run (see Figure S3e). The event of longitudinal bond breaking in the external zigzag atomic row belonging to a hexagon has been also detected. However, such a reaction leads to formation of a one-coordinated atom. Therefore, the bond is recovered in 0.1–1 ps after breaking.

**Estimate of the maximal length of single chains**

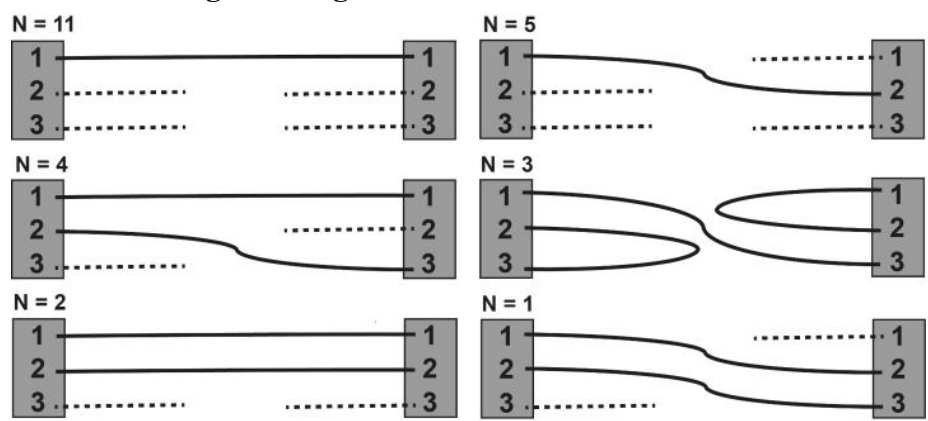

**Figure S4**: Schematic representation of all configurations of attachment of anchored and dangling chains to the fixed atoms at the 3-ZGNR ends generated under heat treatment at temperature 2500 K. The chains attached to the fixed atoms at both ends and only at one end are shown by solid and dotted lines, respectively. Grey rectangles represent the fixed atoms at the ends of the 3-ZGNR, numbers in the rectangles correspond to three zigzag atomic rows. $N$ is the total number of configurations obtained in the simulations for the short and long 3-ZGNRs.

Let us now estimate the maximal length of the single chain that can be achieved under heat treatment of 3-ZGNRs. A sufficiently long 3-ZGNR can be considered as a sequence (along the GNR axis) of short segments where the slow local structure evolution is determined by presence of at least one longitudinal bond broken and long regions with breaking of only cross-bonds, where fast triple chain formation from three zigzag atomic rows of the 3-ZGNR takes place. As mentioned above, the local structure evolution after breaking of at least one longitudinal bond, as a rule, does not lead to the complete rupture of the 3-ZGNR. However, the presence of two or several of such local segments broken can lead to the rupture of 3-ZGNR. The structure rearrangement of local segments with a broken longitudinal bond results in formation of one or two atomic chains passing from one side of the segment to the other (as shown in Figure S3). These chains start from upper (1), middle (2) or lower (3) zigzag atomic rows of the left side and connect them to the same (Figure S3a, S3c and S3d) or different (Figure S3b and S3e) rows on the right side. If the rows with chains attached on the right side of one segment do not have chains on the left side of the next segment, the GNR rupture takes place. For example, the 3-ZGNR rupture occurs if the left segment is structure (b8) and the right one is structure (a3) (see Figure S3). Figure S4 presents all different configurations of attachment of anchored and dangling chains to three zigzag atomic rows (1), (2) and (3) of the fixed atoms at the GNR ends obtained in the simulation runs for the short and long 3-ZGNRs where at least one atomic chain connecting two GNR ends is left. The mirror images are considered as identical. Since diverse configurations with different correspondence between anchored chains and atomic zigzag rows are observed, in a system including many of these configurations, there will be no chain passing through all the junctions and the GNR rupture will take place. Using the probabilities of 19/26 that the segment with a longitudinal cross-bond broken is left with one anchored chain after the structural rearrangement and 7/26 that with two (Figure S4), it can be roughly estimated that the probability of the complete GNR rupture grows from 49% when only two such segments are present to 81% for three, 90% for four and 96% for five. In this simple estimate we assume that the probability for the anchored chains to be attached to any zigzag row is the same, though in reality it is lower for the middle row (Figure S4). On the other hand, even when 6-7 such configurations are present, the probability that there is a chain passing from one GNR end to the other is still of 1-2% and, due to the long regions with only cross-bonds broken separating the segments with broken longitudinal bonds, the chain can be long. Therefore, it seems plausible that the maximal chain length can reach a thousand of atoms.

**Description of structures of dangling and free chains obtained under heat treatment at 2500 K of long and short 3-ZGNR**

If at least one longitudinal bond (between atoms of the same zigzag atomic row of 3-ZGNR) is broken without back formation, not only anchored chains attached to fixed atoms at both ends of 3-ZGNR but also dangling chains attached by one or both ends to fixed atoms only at one end of 3-ZGNR and rarely free chains forms. Several types of these dangling chains are obtained:
- simple chain with one free end, the other end is attached to fixed atoms at the 3-ZGNR end. In the table below simple chain is designated as $Ch_n$, where $n$ is the number of atoms in the chain.
- free simple chain with both free ends. In the table below free simple chain is designated as $Ch^f_n$, where $n$ is the number of atoms in the chain.
- free ring. In the table below free ring is designated as $R^f_n$, where $n$ is the number of atoms in the ring.
- loop: a chain with both ends attached to fixed atoms at the same end of the 3-ZGNR. In the table below simple loop which contains only two-coordinated atoms is designated as $L_n$, where $n$ is the number of atoms in the simple loop. Two types of complex loops which contain one and two three-coordinated atoms are also obtained: a loop with a single one-coordinated atom connected to it (designated as $L_ns_1$ in the table below) and two loops consisting of $n$ and $m$ atoms, connected by a single two-coordinated atom (designated as $L_ns_2L_m$ in the table below);
- lasso: a chain which is attached to fixed atoms at one end of the 3-ZGNR and connected to a
- loop. The loop contains a single three-coordinated atom where the chain is connected to it. In the table below lasso is designated as $LS_{n;m}$, where $n$ is the number of atoms in chain and $m$ is the number of atoms in the loop.

**Table S1.** Calculated characteristics of anchored, dangling and free chains obtained under heat treatment at 2500 K of short and long 3-ZGNRs in all simulation runs where less than three atomic chains attached to fixed atoms at both ends of 3-ZGNR form: simulation run number, transformation time $\tau$ (in ps), number of anchored chains attached to fixed atoms at both ends of 3-ZGNR, panel of Figure 4a with schematic representation of all configurations of anchored and dangling chains attachment to fixed atoms at ends of 3-ZGNR, types and sizes of dangling and free chains

| Simulation run number | Transformation time $\tau$ | Number of anchored chains | panel of Figure 4a | dangling and free chains |
|---|---|---|---|---|
| Short 3-ZGNRs | | | | |
| 1 | 35.7 | 2 | (a5) | $Ch_{15}$, $Ch_{45}$ |
| 2 | 48.0 | 1 | (a1) | $L_{26}$, $L_{92}$ |
| 3 | 61.8 | 1 | (a4) | $L_{68}$, $L_{50}$ |
| 4 | 52.5 | 1 | (a1) | $L_{82}$, $LS_{19;1}$, $L_{16}$ |
| 5 | 48.2 | 1 | (a1) | $R^f_{33}$, $L_{35}$, $L_{50}$ |
| 6 | 17.0 | 1 | (a1) | $L_{34}$, $L_{86}$ |
| 7 | 64.9 | 1 | (a4) | $L_{54}$, $L_{26}$, $L_{40}$ |
| 8 | 236.0 | 1 | (a1) | $Ch_{16}$, $L_{29}s_1$, $Ch_{72}$ |
| 9 | 66.5 | 1 | (a1) | $Ch_{55}$, $Ch_{33}$, $L_{32}$ |
| 10 | 36.0 | 1 | (a4) | $L_{47}$, $L_{71}$ |
| 11 | 24.0 | 1 | (a2) | $L_{107}$, $L_{11}$ |
| 12 | 51.3 | 1 | (a2) | $L_{36}$, $Ch_{10}$, $Ch_{36}$ |
| 13 | 287.9 | 1 | (a1) | $Ch_8$, $L_{110}$ |
| Long 3-ZGNRs | | | | |
| 1 | 117.4 | 2 | (a3) | $Ch_{80}$, $Ch_{110}$ |
| 2 | 153.1 | 2 | (a5) | $Ch_{56}$, $Ch_{134}$ |
| 3 | 130.3 | 2 | (a3) | $Ch_{152}$, $Ch_{39}$ |
| 4 | 191 | 2 | (a3) | $Ch_{26}$, $Ch_{165}$ |
| 5 | 144.2 | 2 | (a6) | $Ch_{70}$, $Ch_{96}$ |
| 6 | 110.1 | 2 | (a3) | $Ch_{113}$, $Ch_{79}$ |
| 7 | 153.3 | 1 | (a2) | $Ch^f_{47}$, $Ch^f_{64}$, $L_{30}$, $L_{160}$, $Ch^f_{82}$ |
| 8 | 159.8 | 1 | (a1) | $Ch_{190}$, $L_{140}$, $L_{50}$ |
| 9 | 107.6 | 1 | (a1) | $L_{144}s_2L_{224}$, $L_{16}$ |
| 10 | 117.2 | 1 | (a1) | $L_{362}$, $L_{22}$ |
| 11 | 142.4 | 1 | (a1) | $L_{123}$, $L_{161}s_2L_{97}$ |
| 12 | 105.7 | 1 | (a2) | $Ch_{182}$, $Ch_{70}$, $Ch_{121}$, $Ch_{10}$ |
| 13 | 119.8 | 1 | (a2) | $Ch_{124}$, $Ch_{31}$, $Ch_{226}$ |
| 14 | 109.6 | 0 | - | $LS_{90;8}$, $LS_{185;12}$, $L_{167}$, $Ch_{16}$, $Ch_{98}$ |